\documentclass[11pt]{article}  
\usepackage{ltexpprt} 
\usepackage{sodafixes}

\usepackage{graphicx}
\usepackage{cite}
\usepackage{url}
\usepackage{amsfonts}
\usepackage{amssymb}

\def\R{\mathbb R}

\def\H{\mathbb H}
\def\lfs{{\mathop{\rm lfs}}}
\def\perim#1{|\partial #1|}

\begin{document}

\title{Squarepants in a Tree:\\
Sum of Subtree Clustering and Hyperbolic Pants Decomposition} 

\author{David Eppstein\thanks{Computer Science Department,
Donald Bren School of Information \& Computer Sciences,
University of California, Irvine; eppstein@uci.edu}}

\date{ }

\maketitle

\pagestyle{empty}
\thispagestyle{empty}

\begin{abstract}
\small\baselineskip=9pt
We provide efficient constant factor approximation algorithms for the problems of finding a hierarchical clustering of a point set in any metric space, minimizing the sum of minimimum spanning tree lengths within each cluster, and in the hyperbolic or Euclidean planes, minimizing the sum of cluster perimeters. Our algorithms for the hyperbolic and Euclidean planes can also be used to provide a {\em pants decomposition}, that is, a set of disjoint simple closed curves partitioning the plane minus the input points into subsets with exactly three boundary components, with approximately minimum total length.
In the Euclidean case, these curves are squares; in the hyperbolic case, they combine our Euclidean square pants decomposition with our tree clustering method for general metric spaces.
\end{abstract}

\section{Introduction}

\begin{figure}[t]
\centering
\includegraphics[width=4in]{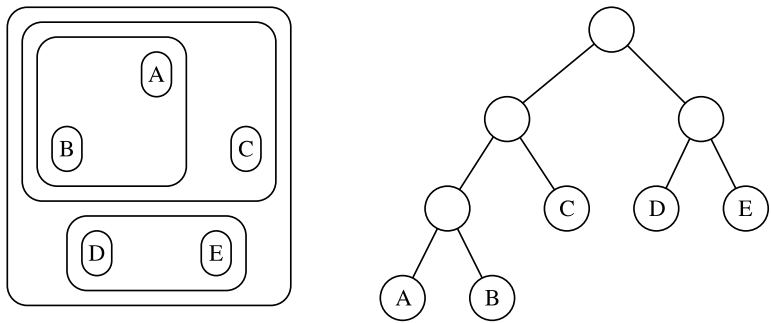}
\caption{A hierarchical clustering of five points (left) and the corresponding binary tree (right).}
\label{fig:hierarchy}
\end{figure}

A {\em hierarchical clustering} of a finite set of points can be visualized as a binary tree, having the points at its leaves. In such a tree, we can form a cluster for each internal node, of the points descending from it. These clusters, together with the empty set, will form a family $\cal F$ of subsets of the points, with the property that any two subsets in the family are either disjoint or related by containment; this family is maximal in the sense that no additional set can be added to it while preserving this property.
Equivalently, a family of maximal sets of this type can be viewed as forming a binary tree, with an internal node per nonempty set; each node has as its children the maximal subsets in the family. Figure~\ref{fig:hierarchy}) shows these two equivalent views of a hierarchy.

There has been much work on heuristics for hierarchical clustering of points in metric spaces~\cite{Joh-PM-67}, often based on {\em agglomerative clustering} methods that start from singleton clusters and repeatedly merge pairs of clusters into single larger clusters until only one set remains~\cite{BorOstRab-ML-04,Epp-JEA-00,Zup-82}. For instance, the {\em single linkage} clustering method is essentially equivalent to Kruskal's algorithm for minimum spanning tree construction, and the {\em neighbor-joining} method~\cite{SaiNei-MBE-87}, which can be defined as a general clustering technique in this way, is widely used for reconstructing evolutionary trees.

However, there has been less work on problems of finding a clustering that optimizes some objective function measuring the overall quality of the clustering~\cite{Das-COLT-02}. In this paper we consider problems of finding a clustering minimizing the sum of cluster sizes, where we may measure the size of a cluster either by the length of its minimum spanning tree (for general metric spaces) or by the perimeter of its convex hull (for the Euclidean and hyperbolic planes). We are unaware of prior work on these versions of the optimal hierarchical clustering problem.

We also consider related problems of cutting the plane by a system of disjoint simple closed curves such that each component of the plane minus the curves and the input points has three boundary components, minimizing the total curve length. Such a {\em shortest pants decomposition} has been approximated in the Euclidean case by Poon and Thite~\cite{cs.CG/0602080}, and related decompositions of surfaces into simple components by short curves have proven useful as building blocks for other topological computations~\cite{EriVer-SODA-06}. For the Euclidean case, we provide a simple quadtree-based approximation algorithm that is less accurate but more efficient than that of Poon and Thite. We also provide similar approximation algorithms for pants decomposition in the hyperbolic plane; to our knowledge this version of the pants decomposition problem has not been studied previously.

This paper is independent of, and concurrent with, a recent paper by Krauthgamer and Lee~\cite{KraLee-FOCS-06}, which claims to be the first to study approximation algorithms in hyperbolic spaces. Krauthgamer and Lee obtain a polynomial time approximation scheme for the traveling salesman problem in any fixed dimensional hyperbolic space, by a technique very similar to our method for hyperbolic clustering, in which points in low-diameter clusters are approximated with Euclidean spaces while the connections between the clusters are approximated by trees.

\section{New Results}

We prove the following results.

\begin{itemize}
\item We formulate the problem of hierarchical clustering minimizing the sum of spanning tree lengths of the clusters, for general metric spaces, show that it is NP-complete, and provide a constant factor approximation algorithm for the problem.
\item We formulate the problem of hierarchical clustering in the Euclidean plane, minimizing the sum of convex hull perimeters, and relate it to the previously studied problem of optimal pants decomposition~\cite{cs.CG/0602080}.  We provide a simple example showing that the two problems do not always have equal solutions, but we show that they can both be approximated to within a constant factor of the optimal total length by a simple quadtree-based clustering algorithm, in time $O(n\log n)$.
\item By analogy with the Euclidean case, we formulate the sum-of-perimeter clustering and pants decomposition problems in the hyperbolic plane. We provide an approximation algorithm based on a combination of our Euclidean technique (for point sets with diameter $O(1)$) and our general metric space technique (for point sets with closest distance $\Omega(1)$). Our hyperbolic approximation uses a lemma that may be of independent interest: for any hyperbolic point set with closest distance $\Omega(1)$, the convex hull and minimum spanning tree have lengths within a constant factor of each other.
\end{itemize}

\section{Hardness of Sum of Subtree Clustering}

\begin{figure}[t]
\centering
\includegraphics[width=4.5in]{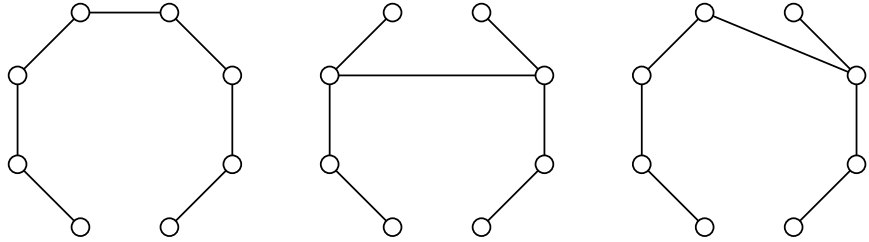}
\caption{The three $3$-bisectable trees.}
\label{fig:3divtrees}
\end{figure}

Define an $i$-bisectable tree, for integer $i\ge 0$, as follows: a $0$-bisectable tree is just a tree with a single vertex, and an $i$-bisectable tree, for integer $i>0$, is formed by connecting any two $(i-1)$-bisectable trees by a single edge connecting any two of their vertices. An $i$-bisectable tree always has exactly $2^i$ vertices. We say that a tree is {\em bisectable} if it is an $i$-bisectable tree for some~$i$. Up to isomorphism of free trees, there is only one $1$-bisectable tree (a single edge), one $2$-bisectable tree (a path of four vertices), and three $3$-bisectable trees (Figure~\ref{fig:3divtrees}).

Any $i$-bisectable tree either has at most a single nontrivial automorphism that exchanges its two $(i-1)$-bisectable subtrees. Based on this observation, if we let $d_i$ represent the number of $i$-bisectable trees, $s_i$ denote the number of symmetric $i$-bisectable trees, and $a_i=d_i-s_i$ the number of asymmetric $i$-bisectable trees (e.g., $d_3=3$, $s_3=2$, and $a_3=1$), we can compute these numbers by the following recurrence.
\begin{eqnarray*}
s_i &=& 2^{i-1}a_{i-1} + 2^{i-2}s_{i-1}\\
a_i &=& {s_i\choose 2}\\
d_i &=& a_i+s_i
\end{eqnarray*}
Using this recurrence we counted $136$ $4$-bisectable trees, and $2098176$ $5$-bisectable trees.

\begin{lemma}
\label{lem:bisectable-recognition}
We can test whether an $n$-node tree is bisectable in time $O(n)$.
\end{lemma}

\begin{proof}
Given a tree $T$, perform the following steps:
\begin{itemize}
\item Choose a root for $T$ arbitrarily.
\item For each vertex $v$ of $T$, count the number of descendants of $v$ (including $v$ itself). This can be done by a single postorder traversal of $T$, as the number of descendants of $v$ is one plus the sum of the numbers of descendants of each of its children.
\item Define an edge of $T$ to be \emph{odd} if the farther of its two endpoints from the tree root has an odd number of descendants. Identify the set of odd edges.
\item If $T$ does not have exactly $n/2$ odd edges then $T$ is not bisectable. Otherwise, form a tree $T'$ with $n/2$ vertices by contracting each odd edge of $T$. $T$ is bisectable if and only if $T'$ is bisectable, which can be tested by a recursive application of the same algorithm.
\end{itemize}
If $T$ is bisectable, consider the sequence of forests of $2^i$ subtrees formed by splitting $T$ according to the first $i$ levels of the bisection hierarchy of $T$. Then it is straightforward to show by induction on $i$ that, at each step of this sequence before the last, the set of edges that are odd in their subtrees remains unchanged. Therefore, if $T$ is bisectable, the odd edges connect pairs of vertices in the forest of $n/2$ subtrees at the bottom level of the hierarchy, and contracting these edges does not change the bisectability of the remaining hierarchy; thus, the algorithm will report correctly that $T$ is bisectable. Conversely if the algorithm reports that $T$ is bisectable, a bisection hierarchy for $T$ can be constructed from the corresponding hierarchy for $T'$. Thus, the algorithm correctly reports the bisectability of $T$.

Each step of the algorithm except for the recursive call takes linear time. The sizes of the subtrees passed as arguments to recursive calls shrink in a geometric series, so the total time for the overall algorithm is also linear.
\end{proof}

\begin{theorem}
\label{thm:divtree-npc}
It is $NP$-complete, given an undirected graph $G$ with $2^i$ vertices, to determine whether
$G$ has an $i$-bisectable subtree.
\end{theorem}

\begin{figure}[t]
\centering
\includegraphics[width=4in]{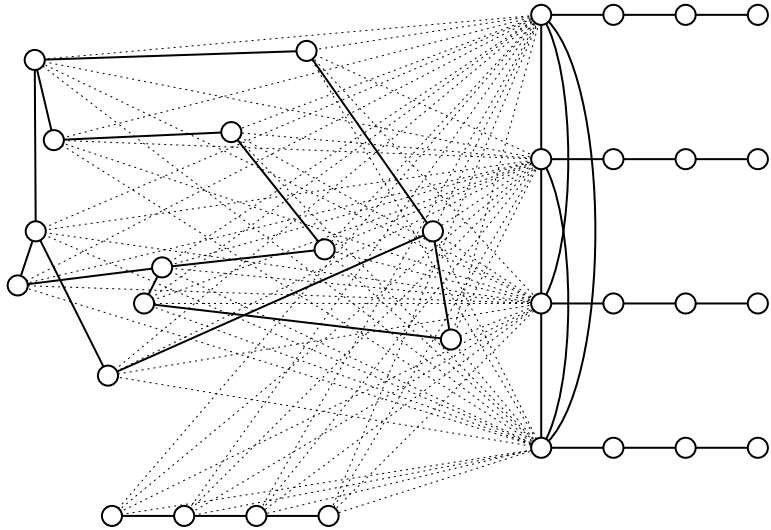}
\caption{Reduction for NP-completeness of finding bisectable subtrees. On the upper left is a twelve-vertex graph $G$ for which we wish to solve the $H$-matching problem; lower left is a path used to pad its size to a power of two. On the right we have the paths $p_i$, with their endpoints $u_i$ connected to each other. The edges between $u_i$ and the vertices on the left are shown light and dashed in order to avoid obscuring the drawing.}
\label{fig:npcredux}
\end{figure}

\begin{proof}
Membership in $NP$ follows since we can use Lemma~\ref{lem:bisectable-recognition} to test whether a given subtree is bisectable.
To show that the problem is $NP$-complete, we reduce from the known $NP$-complete problem of $H$-matching (that is, covering all vertices of a graph by disjoint copies of a fixed subgraph~$H$). $H$-matching is known to be $NP$-hard for all connected subgraphs $H$ with more than two vertices~\cite{GarJoh-79,KirHel-STOC-78}; in our case, we use as $H$ the unique $3$-bisectable tree; that is, the path on four vertices.

So, suppose we are given a graph $G$, and wish to determine whether $G$ has an $H$-matching. We reduce the problem to the existence of a bisectable subtree on a larger graph. We can assume without loss of generality that the number of vertices in $G$ is bisectable by four; otherwise $G$ can have no $H$-matching. Let $G'$ be the disjoint union of $G$ with paths of length four, sufficient so that the number $n$ of vertices in $G'$ is a power of two.
Finally, form graph $G''$ by adding to $G'$ $n$ additional vertices, in the form of $n/4$ paths $p_i$. Choose an endpoint $u_i$ of each such path, and add additional edges between every pair of vertices $u_i,u_j$ and between each $u_i$ and each vertex of $G$. The completed graph $G''$ is depicted in Figure~\ref{fig:npcredux}.

We claim that $G$ has an $H$-matching if and only if $G''$ has an $i$-bisectable subtree.
For, if $G$ has an $H$-matching, then we can cover $G''$ by paths of length eight by connecting each $p_i$ to one of the paths in the $H$-matching; we can then merge these paths into a $i$-bisectable tree using the edges between pairs $u_i,u_j$. Conversely, if $G''$ has an $i$-bisectable subtree,
then by repeatedly partitioning that subtree according to the definition of divisibility, we find a cover of $G''$ by paths of length four. Each path $p_i$ must be in that cover (because it is the only path that covers the endpoint most distant from $u_i$) so the remaining paths must form an $H$-matching of $G$.

Thus we have reduced $H$-matching to our $i$-bisectable subtree problem, completing the proof of $NP$-completeness of that problem.
\end{proof}

The same reduction shows more generally that, given $G$ and $i\ge 2$, it is $NP$-complete to determine whether $G$ can be covered by disjoint $i$-bisectable subtrees.

\begin{theorem}
It is $NP$-complete, given a metric space $\cal M$ (specified as its distance matrix) and a number $K$, to determine whether $\cal M$ has a hierarchical clustering in which the sum of minimum spanning tree lengths of the clusters is at most $K$.
\end{theorem}

\begin{proof}
Membership in $NP$ is straightforward, since we may demonstrate that there exists a clustering with small total size by exhibiting the clustering.

To prove $NP$-hardness, we reduce from the problem of finding an $i$-bisectable subtree, which we have seen is $NP$-complete. Suppose we are given a graph $G$, with $2^i$ vertices, and wish to determine whether $G$ contains an $i$-bisectable subtree. From $G$ we form a metric space $\cal M$ in which the points are the vertices of $G$; two points are at distance one if they are connected by an edge in $G$, and at distance two otherwise. We set $K=i2^i-2^i+1$.

If $G$ contains an $i$-bisectable subtree $T$, we form a clustering by recursively splitting $T$ into smaller bisectable subtrees. Each cluster in this clustering has a spanning tree with unit length edges, namely the subtree of $T$ induced by its vertices, so we can calculate that the total length of all minimum spanning trees of clusters is exactly $K$. Conversely, any clustering of length $K$ must have $2^{i-j}$ vertices in each cluster at level $j$ of the clustering, or else even if all spanning tree edges have unit length the total length would exceed $K$. Further, each pair of clusters at level $j$ must be connected by at least one edge of $G$, or else some spanning tree would have an edge of length two, again causing the total length to exceed $K$. So, in this case, choosing an edge connecting each pair of clusters at each level of the tree gives us an $i$-bisectable subtree of $G$.

Thus we have reduced the $i$-bisectable subtree problem to our clustering problem, completing the proof of $NP$-completeness of that problem.
\end{proof}

\section{Approximate Sum of Subtree Clustering}

Suppose that we wish to find a hierarchical clustering of a set of points in a metric space, approximately minimizing the sum of cluster sizes, where we measure the size of a cluster by the total length of the edges in its minimum spanning tree. 

We observe the optimal clustering may have a structure quite different from that of the minimum spanning tree of $\cal M$. For one thing, our NP-completeness reduction shows that, if one forms a graph connecting points at close distances to each other, it may be more important to find a bisectable spanning tree in this graph than a graph of minimum possible weight. For another, even in a tree metric, if we restrict ourselves to partitions formed by recursively splitting the tree on its edges we may form a clustering very far from optimal; for instance, for the star $K_{1,n-1}$ (with unit edge lengths) recursive splitting produces a clustering with total weight $\Omega(n^2)$ while the optimal clustering has total weight $O(n\log n)$.

\begin{figure}[p]
\centering
\includegraphics[width=5in]{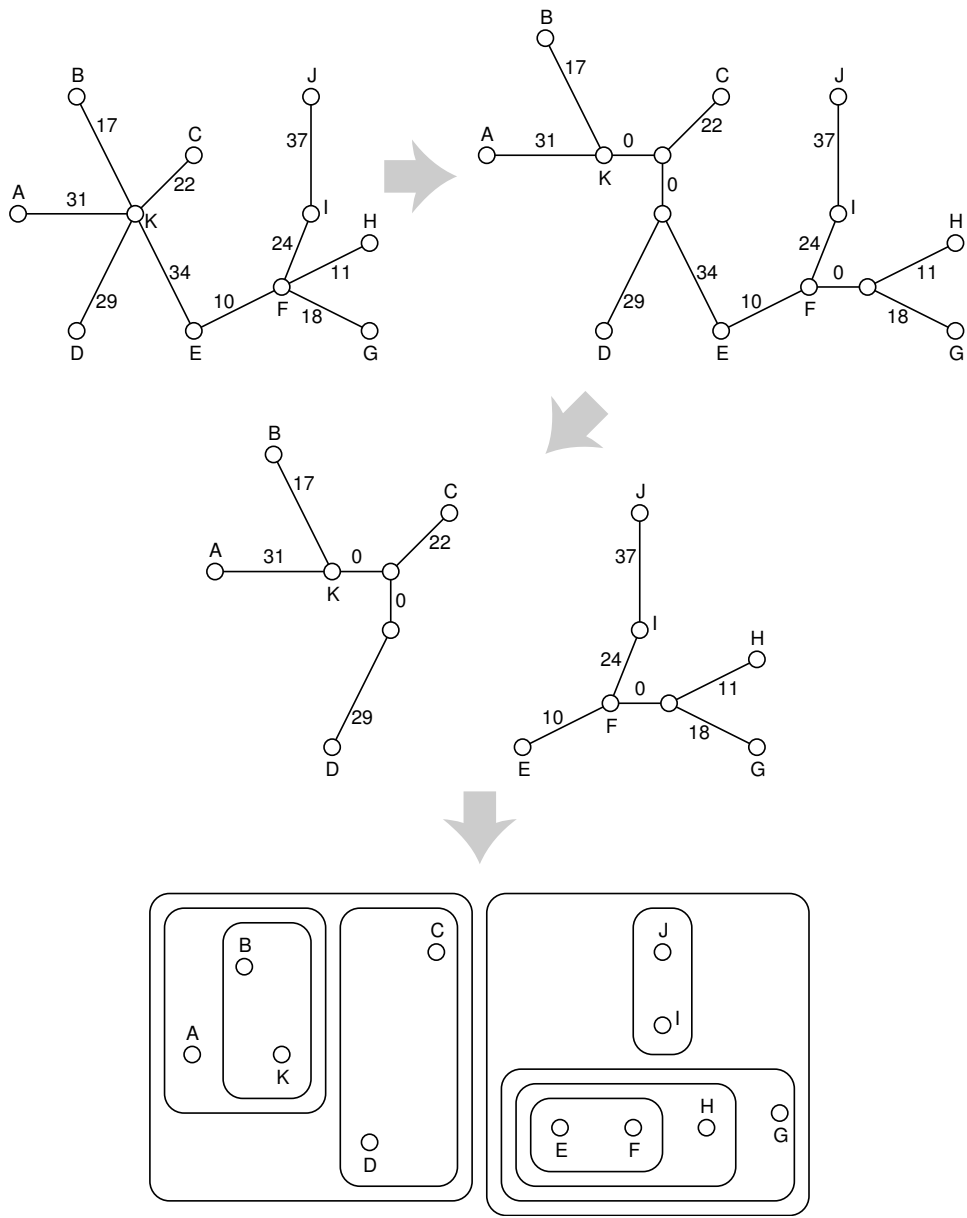}
\caption{Approximation algorithm for subtree clustering. Top left: a minimum spanning tree $T$ of a point set, with edges labeled by length. Top right: tree $T^*$ subdivided from $T$ so that each vertex has degree at most three; new vertices are unlabeled. Center: partitioning $T^*$ to minimize the maximum length of the two resulting subtrees. Bottom: the clustering formed by continuing the partition recursively in each subtree.}
\label{fig:treeapprox}
\end{figure}

Nevertheless, as we now show, the idea of recursively splitting the minimum spanning tree can lead to an efficient approximation for the optimal clustering.
We approximately cluster $\cal M$ by the following algorithm, the operation of which is depicted in Figure~\ref{fig:treeapprox}.

\begin{enumerate}
\item Compute a minimum spanning tree $T$ of the points.
\item While $T$ contains a vertex $v$ with degree greater than three, split $v$ into two vertices connected by an edge of length zero, each adjacent to two or more of the neighbors of $v$. Give one of these two vertices the identity of the original input point it came from.
\item Call the resulting tree resulting from these split operations $T^*$; it has at most three edges per vertex.
\item Find the edge $e$ of $T^*$ such that the maximum total length among the two subtrees $T^*_{\ell}$ and $T^*_{r}$ is as small as possible.
\item Form two clusters consisting of the input points corresponding to vertices in these two subtrees.
\item Use the two subtrees $T^*_{\ell}$ and $T^*_{r}$  to partition these two clusters recursively.
\end{enumerate}

To prove that this algorithm produces an approximation to the optimal clustering, we prove an upper bound on the total weight of the clustering produced by the algorithm, and a matching lower bound on any clustering. The upper bound depends on a standard lemma on tree separators.

\begin{lemma}
\label{lem:treesep}
In any tree with nonnegative edge weights, let $e$ be an edge the removal of which minimizes the maximum weight among the two resulting subtrees. Then the weight of each subtree formed by the removal of $e$ is at most $2/3$ of the total weight of the initial tree.
\end{lemma}

\begin{proof}
Let the tree be $T$ and its weight be $W$.
If $e$ is an edge separating $T$ into subtrees one of which has weight larger than $(2/3)W$, let $f$
be the edge adjacent to $e$ in this heavy subtree, such that the weight of the subtree containing $f$
is larger than the weight of the other subtree adjacent to the same vertex of $e$.
Then the subtree containing $f$ has weight at least $W/3$, so $f$ forms a better split: it partitions the tree into two subtrees, one of which (the one containing $e$) has weight at most $(2/3)W$, and the other of which either has less weight or fewer edges than the heavy subtree for $e$. We can not find an infinite sequence of better splits, so there must be an edge forming a good split as described by the lemma.
\end{proof}

\begin{lemma}
\label{lem:entropy-ub}
Let the edges of the minimum spanning tree $T$ have weights $w_0$, $w_1$, $\ldots$ in descending order, and let $W=\sum w_i$. Then the total weight of the spanning trees of the clusters produced by our clustering algorithm is at most $\sum w_i (1+\log_{3/2}(W/w_i))$.
\end{lemma}

\begin{proof}
An edge $e_i$ of $T$ with weight $w_i$ participates in the minimum spanning trees of clusters at levels $0$, $1$, $2$, $\ldots$ of the clustering, until reaching a level in which its two endpoints are in different clusters. At level $k$, the total weight of the spanning tree in which it participates is (by Lemma~\ref{lem:treesep}) at most $W(2/3)^k$; if $k>\log_{3/2}(W/w_i)$, this total weight is less than $w_i$, so $e_i$ can only participate in trees in at most $\log_{3/2}(W/w_i)$ levels of the hierarchy.
\end{proof}

\begin{figure}[t]
\centering
\includegraphics[width=5.5in]{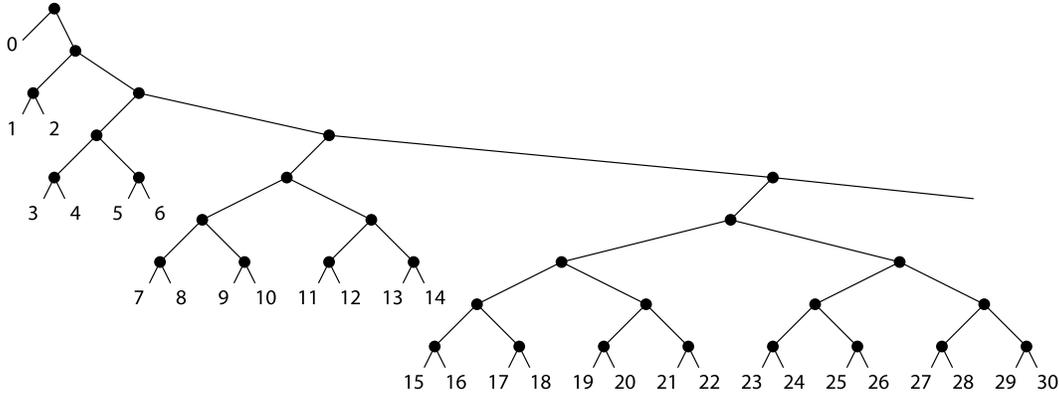}
\caption{A binary code with expected path length $1+\sum 2 w_i\lfloor \log_2(i+1) \rfloor$.}
\label{fig:ranktree}
\end{figure}

\begin{lemma}
\label{lem:von-Neumann-Shannon}
For any decreasing sequence of values $w_i$ with sum $W$,
$$\sum w_i(\frac12+\lfloor \log_2(i+1) \rfloor) \ge \sum \frac12 w_i \log_2(W/w_i).$$
\end{lemma}

\begin{proof}
By scaling the $w_i$ by the same factor, we may assume without loss of generality that $W=1$. We interpret the $w_i$ as probabilities of drawing symbol $i$ in a random variable that takes the indices $0,1,\ldots n-1$ as its values, with index $i$ having probability $w_i$. There exists a binary code (that is, a binary tree having these indices at its leaves) in which the expected path length to a randomly drawn index is $1+\sum 2 w_i\lfloor \log_2(i+1) \rfloor$ (Figure~\ref{fig:ranktree}). The result then follows from Shannon's entropy lower bound of $\sum w_i \log_2(1/w_i)$ on the average path length of any binary code~\cite{Sha-BSTJ-48}.
\end{proof}

The relation between this sum of logs of ranks and entropy is closely related to the efficiency of one-to-one codes in coding theory~\cite{BluPri-TIC-96,LeuCov-TIC-78}; codes similar to the one shown in Figure~\ref{fig:ranktree} have also been used as part of more complex data compression schemes~\cite{BenSleTar-CACM-86,Eli-TIC-87}.
We note that an inequality in the other direction,
$\sum w_i\lfloor \log_2(i+1) \rfloor \le \sum w_i \log_2(W/w_i)$,
follows more trivially: $i+1\le W/w_i$, since otherwise the sum of the first $i+1$ weights would exceed $W$~\cite{Wyn-IC-72}.

\begin{lemma}
\label{lem:entropy-lb}
Let $C$ be any hierarchical clustering in a metric space $\cal M$, let
the edges of the minimum spanning tree of $\cal M$ have weights $w_0$, $w_1$, $\ldots$ in descending order, and let $W=\sum w_i$. Then the total weight of the minimum spanning trees of the clusters in $C$ is at least $\sum \frac12 w_i(1+ \log_2(W/w_i))$.
\end{lemma}

\begin{proof}
Let $C_i$ be the total length of the minimum spanning trees of the clusters at level $i$ of the clustering, and let $F_i$ be the total length of the minimum forest in $\cal M$ having $2^i$ trees.
Note that this minimum forest can be computed by removing the $2^i-1$ largest edges from the minimum spanning tree of $\cal M$.
Then
\begin{eqnarray*}
\sum C_i & \ge & \sum F_i \\
 & = & \sum_i\sum_{j\ge 2^i-1} w_j \\
 & = & \sum w_i(1+\lfloor \log_2(i+1) \rfloor) \\
 & \ge & \sum \frac12 w_i (1+\log_2(W/w_i)), \\
\end{eqnarray*}
where the final inequality in this sequence is Lemma~\ref{lem:von-Neumann-Shannon}.
\end{proof}

\begin{theorem}
\label{thm:metric}
If we are given a minimum spanning tree of a metric space $\cal M$, we may compute in $O(n\log n)$ time a hierarchical clustering, such that the sum of minimum spanning tree lengths of clusters in our clustering is within a factor of $2\log_{3/2}2\sim 3.42$ of the optimal clustering.
\end{theorem}

\begin{proof}
We may implement the approximation algorithm described above by using the dynamic tree median data structure of Alstrup et al.~\cite{AlsHolTho-SWAT-00} to find the edge to be removed at each step of the recursive partition of the tree $T^*$.  This data structure takes time $O(\log n)$ per step, and all other operations of the algorithm may easily be implemented in total time $O(n)$, so the total time for the algorithm is $O(n\log n)$. The approximation ratio for our algorithm follows from Lemmas~\ref{lem:entropy-ub} and~\ref{lem:entropy-lb}.
\end{proof}

\section{Euclidean Sites}

\begin{figure}[t]
\centering
\includegraphics[width=2.5in]{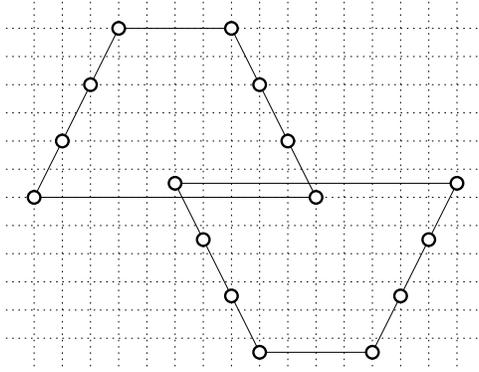}
\caption{A set of 16 points such that the clustering minimizing the total perimeter of the hulls of the clusters has non-disjoint hulls of clusters. The two top-level clusters in the optimal clustering are shown.}
\label{fig:cnp}
\end{figure}

In this section we consider two problems. The first is a clustering problem very similar to the one we have already studied for general metric spaces: hierarchically clustering point sites in the Euclidean plane $\R^2$ in such a way as to minimize the sum of cluster perimeters.
The second problem we consider is one of optimal {\em pants decomposition}. A {\em pair of pants} is a topological surface in the form of a disk with two holes cut into it; that is, it can be represented as a connected subset of the plane the boundary of which has three connected components. A {\em pants decomposition} is a partition of a topological surface into pairs of pants. The Euclidean pants decomposition problem considers as input a set of point sites in the plane, and asks for a family of disjoint closed curves such that removing the curves and the sites from the plane decomposes it into connected components all of which are pairs of pants. Each pair of pants must have one curve as its outer boundary, except for one infinite pair of pants having a boundary at infinity. However the two inner boundaries of each pair of pants may either be other curves or input sites.
Figure~\ref{fig:hierarchy} (left) and Figure~\ref{fig:treeapprox} (bottom) depict points in the plane of the drawing, surrounded by curves; if one removes the outer curve from each of these figures, they can be interpreted as pants decompositions.

The question of minimum-length pants decomposition was proposed by Jeff Erickson and Kim Whittlesey (personal communication) and first attacked algorithmically by Poon and Thite~\cite{cs.CG/0602080}, who provided both a polynomial time approximation scheme and a polynomial time exact algorithm for a restricted version of the problem in which the curves must be rectangles.

\subsection{Clustering versus Pants}

Clustering and pants decomposition are closely related.
We can form a clustering by creating a cluster for the points within each curve of a pants decomposition, and these clusters (together with singleton clusters for each site and a cluster for all sites) form a hierarchical clustering. The length of the curve surrounding each clustering is at least the perimeter of the convex hull of the cluster.
Therefore, the length of a pants decomposition, plus the length of the convex hull of all the sites, is lower bounded by the length of the clustering that minimizes the sum of perimeters.
However, the two problems may have total lengths that differ by arbitrarily large factors, due to the inclusion of the whole set in the clustering and not in the pants decomposition; for instance, the three points $(0,0)$, $(0,1)$, and $(L,0)$ (for large $L$) have a pants decomposition with length $2+\epsilon$ for any $\epsilon$ (formed by a curve around the two closest point) while its optimal clustering has length $2L(1+o(1))$.

Even ignoring the presence or absence of a cluster for the entire point set, the two problems can differ in other ways. Figure~\ref{fig:cnp} depicts a set of 16 points for which the unique optimal clustering has two clusters with overlapping convex hulls. We computed the optimal clustering using a dynamic programming algorithm that determines the optimal clustering for each set of points, by considering all partitions of that set into two previously clustered subsets, in total time $O(3^n)$.

We do not know whether the optimal pants decomposition of these points has the same hierarchical structure, but it must differ in total length. We can also show that there exist point sets for which the optimal clustering and the optimal pants decomposition differ in hierarchy: either the depicted point set is such an example, or we can find such an example by scaling the vertical coordinates of this point set by a factor $0<s<1$ while leaving the horizontal coordinates unchanged. As we scale the points, for sufficiently small $s$ the points become sufficiently close to colinear that their optimal clustering has disjoint hulls; let $s_0$ be the largest $s$ for which this is true. Then scaling by $s_0$ produces a point set in which two clusterings, one with disjoint hulls and one with non-disjoint hulls, have equal lengths, however the pants decomposition corresponding to the clustering with disjoint hulls is strictly shorter than the pants decomposition  corresponding to the clustering with non-disjoint hulls. Scaling by $s_0+\epsilon$ for small epsilon can therefore be seen to have a different hierarchy in its optimal clustering and its optimal pants decomposition. The non-disjoint hulls in this example also imply that it may be difficult to find a polynomial time approximation scheme for the optimal clustering problem, as the technique used by Poon and Thite to approximate the optimal pants decomposition relies on the disjointness of the curves it finds.

\subsection{Euclidean Clustering}

In this section we describe a fast and simple quadtree-based heuristic for finding an approximation to the Euclidean clustering problem. We will later show how to adapt the same algorithm to find similar approximations for pants decomposition. Our algorithm will turn out to be an important subroutine in our later approximation algorithm for similar problems in the hyperbolic plane.

\begin{figure}[t]
\centering
\includegraphics[width=2.75in]{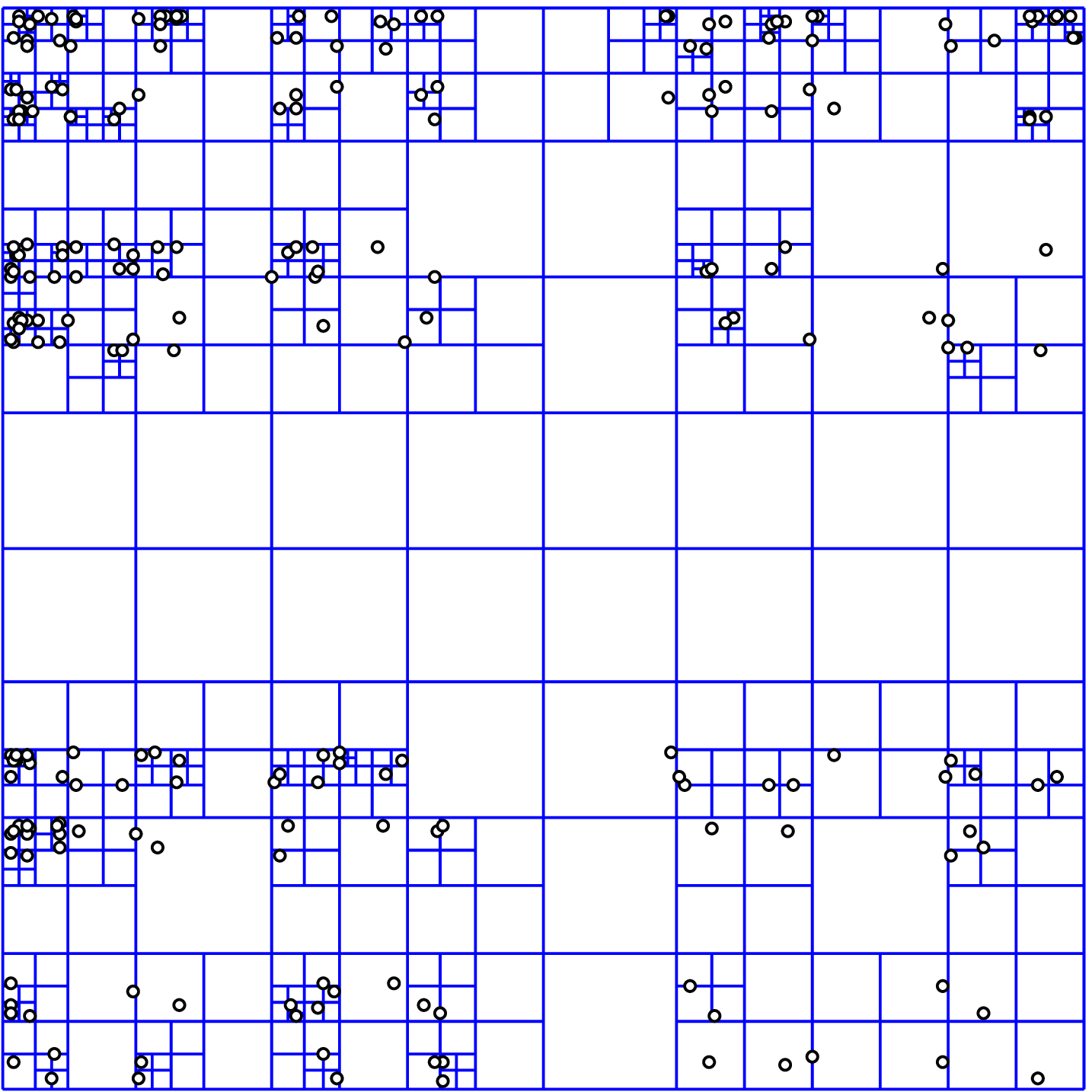}
\qquad
\includegraphics[width=2.75in]{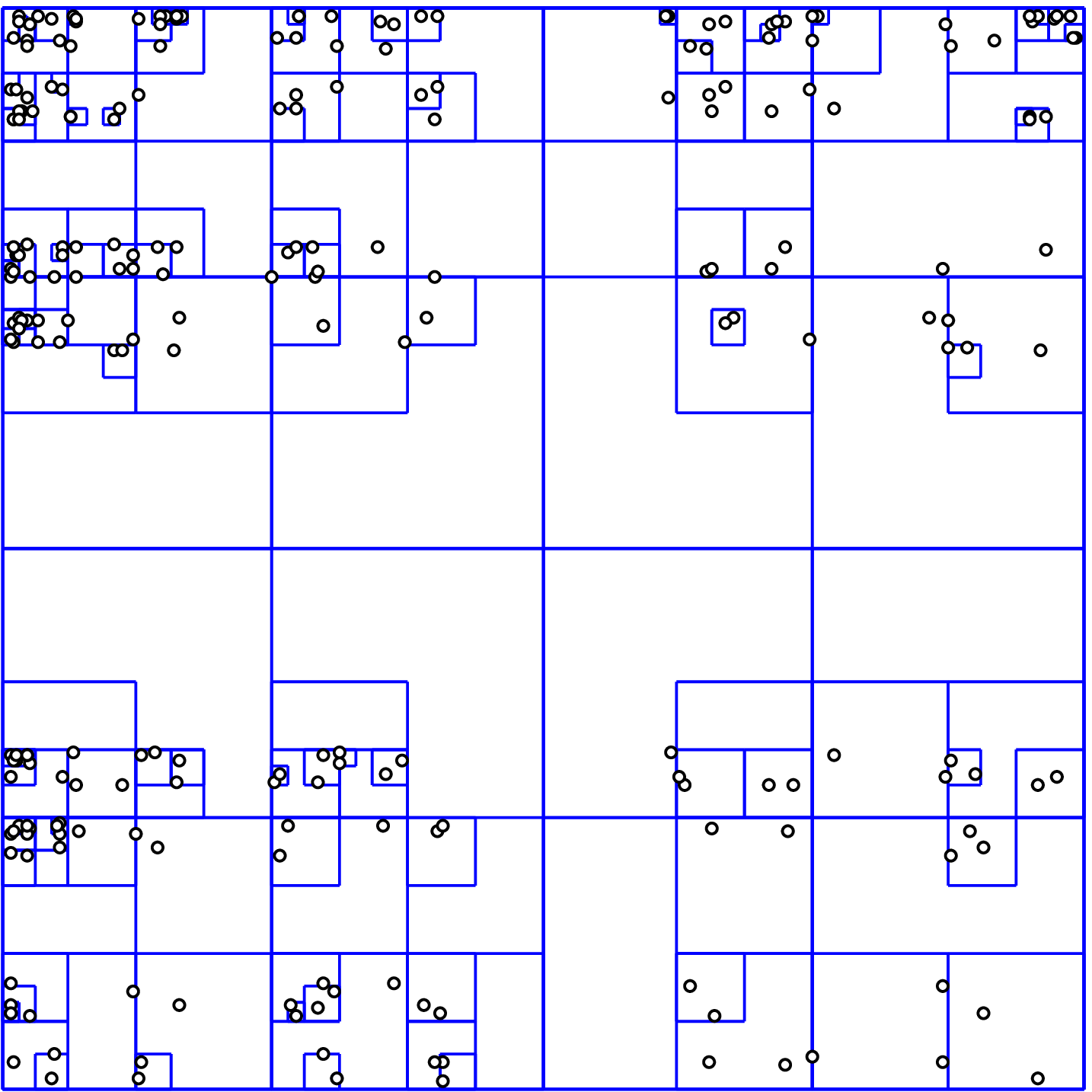}
\caption{A quadtree and the corresponding compressed quadtree.}
\label{fig:cqtree}
\end{figure}

Our fast Euclidean approximation technique is based on the {\em compressed quadtree}~\cite{AluSev-FSTTCS-99,Ber-IPL-93,BerEppTen-WADS-93,Cla-FOCS-83,EppGooSun-SCG-05}.
A quadtree is a well-known recursive space partition data structure formed from a set of sites by surrounding the sites by a square bounding box, then, as long as some minimal square of the structure contains more than one input site, splitting those squares into four smaller squares. To handle cases of ambiguity, we consider a point on the boundary of a square to be outside the square if it is on the lower or right sides of the square, and inside the square otherwise; in this way each square is exactly the disjoint union of its four quadrants. A compressed quadtree is a subset of the squares of a quadtree, consisting only of those squares that contain sites in more than one of their four quadrants.  Figure~\ref{fig:cqtree} depicts a quadtree and compressed quadtree for a set of sites. Each square of a compressed quadtree other than the initial root square is contained in a unique minimal larger square of the compressed quadtree, which we consider to be its parent. We also consider each site to be a node in the structure, with parent the minimal square containing it. Thus, the structure forms a tree, in which the leaves are the input sites, and each internal node corresponds to a square with at least two children, at most one child per quadrant of the square. Since each square has two or more children, there are at most $n-1$ squares in the compressed quadtree. A compressed quadtree for a set of $n$ sites may be built or maintained dynamically in $O(n\log n)$ time, in a model of computation in which we may perform arithmetic and bitwise Boolean operations on the binary representations of the site coordinates~\cite{BerEppTen-WADS-93,EppGooSun-SCG-05}.

To form a hierarchical clustering from a compressed quadtree, we form a cluster for the set of sites inside each square of the quadtree. This is not itself a hierarchical clustering, because some squares in the quadtree may have more than two children. If a square has three children, we also form a cluster combining the sites in two of its children, choosing two out of the three children whose quadrants share a side with each other. If a square has four children, we form two clusters combining the children in adjacent pairs.
As the set of sites inside a square has a convex hull perimeter bounded above by the perimeter of the square, the total convex hull perimeters of the clusters in the resulting hierarchical clustering is bounded above by the total perimeters of the squares in the compressed quadtree.
Note that, in this clustering, disjoint clusters have disjoint convex hulls.

We analyze this clustering using an integral involving {\em local feature size}~\cite{AmeBerEpp-GMIP-98,Rup-Algs-95}, reminiscent of Ruppert's similar analysis of the optimal number of triangles in a bounded-aspect-ratio triangulation~\cite{Rup-Algs-95}. Given a set $S$ of input sites, define the local feature size $\lfs(x,y)$ to be the distance from point $(x,y)$ to the second nearest point in $S$. Let $\square$ denote the root square of our quadtree, which we assume to be a minimal bounding square for the input points; thus, for points $(x,y)\in\square$, $\lfs(x,y)$ is bounded above by the diagonal length of the square and below by half the minimal separation between any two points.

\begin{lemma}
The total perimeter of the clusters in our clustering is $O(\int_{(x,y)\in \square} 1/\lfs(x,y)\,dx\,dy)$.
\end{lemma}

\begin{proof}
We prove more generally that, in the (non-compressed) quadtree for $S$, the sum of perimeters of the squares $C_i$ is $O(\int_{(x,y)\in \square }1/\lfs(x,y)\,dx\,dy)$. To do this, consider the following charging scheme:
each square $C_i$ is initially allocated a charge, equal to its perimeter $\perim{C_i}$. Then, in order from larger squares to smaller ones, we reallocate this charge by the following process: if square $C_i$ has all four quadrants containing two or more input sites, we remove the charge from $C_i$ and partition it equally among its four children.  In this way, the total charge remains unchanged, while each square may receive charge at most equal to $\perim{C_i}/2^j$ from its $j$th ancestor, so the total charge is at most $2\perim{C_i}$.

Next, in any square $C_i$ with nonzero charge containing two or more sites, we reallocate the charge from $C$ to a child $C_j$ of $C$ that contains zero or one sites.  This child could not have been charged by its parent in the previous reallocation step, and so ends up with a total charge at most equal to $5\perim{C_j}$.

Finally, we observe that all charge has been concentrated in squares with zero or one sites, and that each such square has a charge proportional to its perimeter. These squares are disjoint, and within any such square, the charge is at most proportional to the integral of $1/\lfs(x,y)$.
\end{proof}

\begin{lemma}
\label{lem:E-lb}
In any hierarchical clustering, the sum of perimeters of convex hulls of clusters is $\Omega(\int_{(x,y)\in\square} 1/\lfs(x,y)\,dx\,dy)$.
\end{lemma}

\begin{proof}
Let $S_i$ be the clusters in some clustering of $S=S_0$.
For each cluster $S_i$ in the clustering, let $C_i$ be a minimal bounding square, let $3C_i$ denote a concentric square three times as wide, and let $D_i=(3C_j\setminus 3C_i)\cap\square$, where $S_j$ is the parent of $S_i$ in the clustering. Further, let $L(x,y,i)=1/d((x,y),S_i)$ where $d$ denotes the Euclidean distance to the closest point in cluster $S_i$. Then
\begin{eqnarray*}
\sum_i \perim{S_i}
&\ge&\frac12 \sum_i \perim{C_i}\\
&\ge&\frac16 \sum_i \perim{3C_i}\\
&\ge&\frac1{12} \sum_i \perim{D_i},
\end{eqnarray*}
because each bounding square has at most twice the perimeter of its cluster, and in the final sum of this sequence of inequalities, the square $3C_i$ contributes to the boundary of two clusters $D_j$ for its two children.

Further, for any cluster $S_i$,
$$\perim{D_i}=\Omega(\int_{D_i} L(x,y,i)\,dx\,dy).$$
To show this, form a sequence of concentric squares, starting at $C_i$ and doubling in size at each step of the sequence until the last member of the sequence contains all of $D_i$.
Within any of the annular regions between two adjacent pairs of squares in this sequence, the integrand is inversely proportional to the perimeters of the two squares, so the integral in this same annular region is directly proportional to its outer square's perimeter. The overall integral is the sum of the integrals within each of these annular regions, a sum that is proportional to a geometric series adding to the perimeter of $D_i$.

Finally, observe that, for $(x,y)\in\square$,
$$1/\lfs(x,y)\le L(x,y,0)\le\max_{(x,y)\in D_i} L(x,y,i).$$
The first equality holds because the local feature size is a distance to some point $p$ in $S=S_0$, and the second holds because the clusters containing $p$ correspond to sets $D_i$ that together cover $\square$.

Putting these claims together,
\begin{eqnarray*}
\sum_i \perim{S_i}
&=&\Omega( \sum_i \perim{D_i} )\\
&=&\Omega( \sum_i \int_{D_i} L(x,y,i)\,dx\,dy )\\
&=&\Omega( \int_{\square} \sum_{(x,y)\in D_i} L(x,y,i)\,dx\,dy )\\
&=&\Omega( \int_{\square} \max_{(x,y)\in D_i} L(x,y,i)\,dx\,dy )\\
&=&\Omega( \int_{\square} 1/\lfs(x,y)\,dx\,dy ),
\end{eqnarray*}
as was to be shown.
\end{proof}

We note that a similar argument based on local feature sizes can be used to provide an alternative proof for our previous result~\cite{Epp-DCG-94} that quadtree triangulations have total length within a constant factor of the minimum length Steiner triangulation of a point set.

Putting these results together, we have:

\begin{theorem}
\label{thm:eclust}
In $O(n\log n)$ time, we can construct a hierarchical clustering with total cluster perimeter within a constant factor of optimal for any set of Euclidean points.
\end{theorem}

\subsection{Euclidean Pants}

We now show how to convert the hierarchical clustering resulting from the method of Theorem~\ref{thm:eclust} into a pants decomposition that approximates the optimal pants decomposition.
We must be careful in doing this, as (due to the pants decomposition lacking a curve that surrounds the entire point set) the optimal pants decomposition may itself have significantly lower length than the compressed quadtree. Nevertheless, we show that the quadtree based clustering leads to a good pants decomposition.

\begin{theorem}
\label{thm:epants}
In $O(n\log n)$ time, we can construct a pants decomposition with total length within a constant factor of optimal for any set of Euclidean points.
\end{theorem}

\begin{proof}
Recall that  Theorem~\ref{thm:eclust} produces a clustering in which disjoint clusters have disjoint convex hulls.
Let $\epsilon$ be $n^{-2}$ times the minimum distance between hulls of disjoint clusters in this clustering; this distance may be found in $O(n\log n)$ time via a medial axis computation. For each cluster $C_i$ in the clustering produced by Theorem~\ref{thm:eclust} other than the cluster of all sites, surround $C_i$ by a curve at distance $|C_i|\epsilon$ from the convex hull of $C_i$. 
The total length of this curve is $2\pi |C_i|\epsilon$ plus the length of the convex hull; these $2\pi |C_i|\epsilon$ terms are negligable, even when added up over all the clusters, and may be ignored for the rest of our analysis. Finally, when pairing up quadrants of the root square to form the overall hierarchical clustering, do so in the way that minimizes the overall length.

We distinguish two cases in our analysis. Let $s$ be the side length of the minimum enclosing square used as the root of our quadtree algorithm. In the first case, the optimal clustering includes a cluster with diameter at least $s/2$. In this case, we may add another curve surrounding the entire point set without increasing the total length of the optimal decomposition by more than a constant factor. As in Theorem~\ref{thm:eclust}, the total length of this augmented optimal decomposition is proportional to that of the decomposition found by our quadtree algorithm.

In the second case, the optimal clustering includes two subsets of small diameter, separated by a larger distance. Let $B_1$ and $B_2$ be the minimum bounding squares of these two subsets.
Then $B_1$ and $B_2$ must touch opposite sides of the quadtree's root square; since they each have small diameter, no quadrant of the root square can contain points from both clusters. Therefore, all curves found by our pants decomposition are contained within (a small dilation of) $B_1$ or $B_2$.
We can then follow the same analysis used in the proof of Theorem~\ref{thm:eclust} to show that
the length of the quadtree-based pants decomposition is upper bounded by
$\int_{B_1\cup B_2} 1/\lfs(x,y)\,dx\,dy$ while the length of the optimal pants decomposition is lower bounded by the same quantity.
\end{proof}

\section{Hyperbolic Pants}

We now describe analogous problems of clustering minimizing the sum of convex hull perimeters, and of optimal pants decomposition, for the hyperbolic plane instead of the Euclidean plane.  Point sets in the hyperbolic plane with constant diameter can be well approximated by Euclidean point sets, so in this case we could apply our quadtree-based Euclidean approximation, but this approach does not work for more widely spaced point sets.  Our eventual algorithm
combines our Euclidean square pants decomposition with our tree clustering method for general metric spaces.

For simplicity of exposition we describe our algorithms as based on primitives that can compute exact distance-based predicates of points in hyperbolic space, ignoring questions of how such points are represented. However, as our eventual result is an approximation algorithm, our results can be extended without difficulty to a computational model in which all distance computations are approximate; we omit the details.

The connection between tree clustering (in which we attempt to optimize the sum of spanning tree lengths) and the hyperbolic clustering problem (in which we instead attempt to optimize the sum of convex hull perimeters) can be made more concrete by the following lemma.

\begin{figure}[p]
\centering
\includegraphics[width=6in]{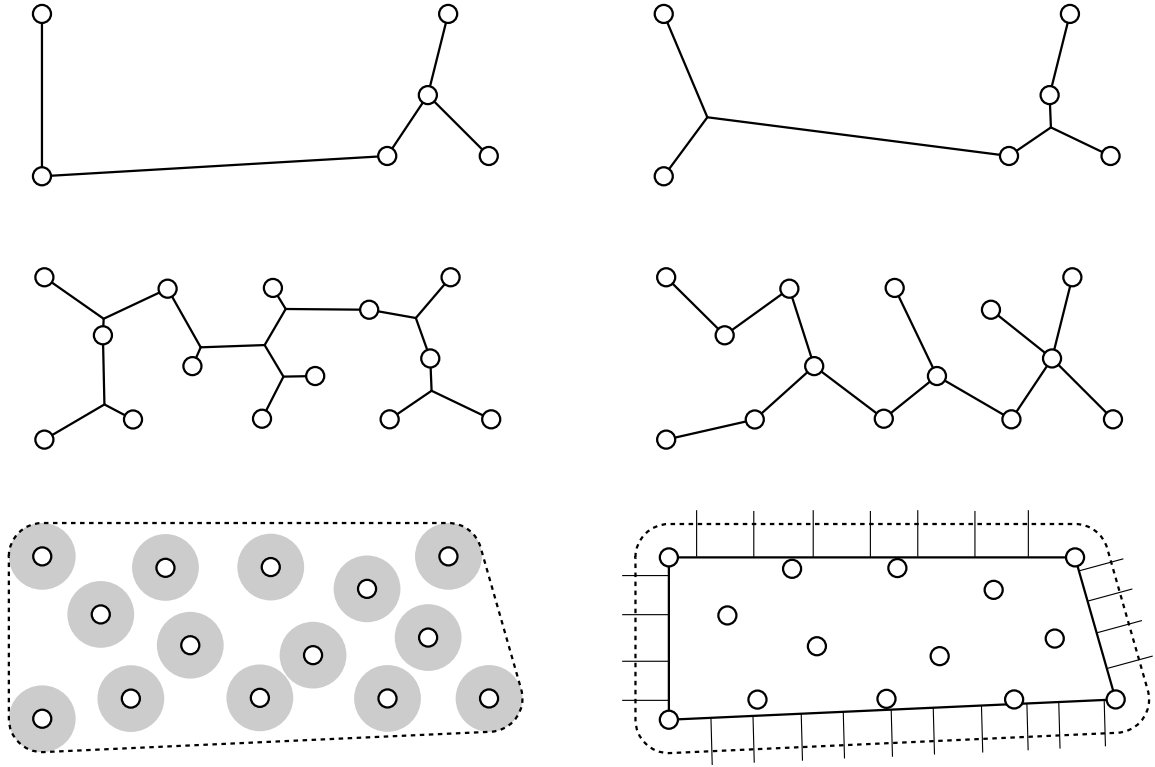}
\caption{Steps in the proof of Lemma~\ref{lem:msthull}. The minimum spanning tree $T$ of $S$ (top left) is proportional in length to its minimum Steiner tree (top right). Augmenting $S$ to a maximal $\epsilon$-separated set $S'$ does not decrease the Steiner tree length (center left), which in turn is less than the length of the minimum spanning tree $T'$ of $S'$ (center right). The $|S'|-1$ edges of $T'$, each of length $\Theta(1)$, have a total length proportional to the total area of a collection of radius-$\epsilon/2$ disjoint disks around each point of $S'$ (bottom left), which in turn is bounded by the area of a set $C$ formed by expanding the hull of $S$ by distance $\epsilon/2$. By hyperbolic isoperimetry, the area of $C$ is bounded by its perimeter, and by partitioning the perimeter of the convex hull $H$ of $S$ into curves of length $\Theta(1)$ and assigning each point of $C$ to the nearest curve we may show that the perimeter of $C$ is proportional to that of $H$ (bottom right). The trees depicted are for illustrative purposes only and may not be the actual optima, as in any case the objects in the figure should be interpreted as belonging to the hyperbolic plane rather than the Euclidean plane.}
\label{fig:msthull}
\end{figure}

\begin{lemma}
\label{lem:msthull}
Let $\epsilon>0$ be a fixed constant, and let $S$ be a set of points in the hyperbolic plane $\H^2$,
such that no two points in $S$ are closer than distance $\epsilon$ to each other.
Then the convex hull perimeter and minimum spanning tree length of $S$ are within constant factors of each other.
\end{lemma}

\begin{proof}
Let $T$ denote the minimum spanning tree of $S$ (Figure~\ref{fig:msthull}, top left) and $H$ denote the convex hull.
In one direction, the vertices of $H$ form a subsequence of an Euler tour of $T$, so (regardless of point separation) $|H|\le 2|T|$.

In the other direction, $T$ of $S$ has length within a constant factor of the minimum Steiner tree of $S$ (Figure~\ref{fig:msthull}, top right), e.g., by a similar Euler tour argument. If we let $S'$ be a maximal set of points having the same convex hull as $S$ and having no pair closer than $\epsilon$, and $T'$ be a minimum spanning tree of $S'$, then the length of $T'$ is again at least a constant factor times that of $T$, as adding points can not decrease the Steiner tree length (Figure~\ref{fig:msthull}, center left) and the minimum spanning tree of $T'$ has length at least that of its Steiner tree (Figure~\ref{fig:msthull}, center right). Note that all edges of $T'$ have length at least $\epsilon$ and at most $2\epsilon$, as if there were any longer edge we could add its midpoint to $S'$ contradicting the assumption that $S'$ is maximal. Thus, the total length of $T'$ (and therefore also that of $T$) is~$O(|S'|)$.

Now, form a disk of radius $\epsilon/2$ around each point of $S'$. These disks are disjoint, and all lie within the set $C$ consisting of the points within distance $\epsilon/2$ of the convex hull of~$S$ (Figure~\ref{fig:msthull}, bottom left).
Therefore, the area of $C$ is $\Omega(|S'|)$, and by a hyperbolic isoperimetric theorem~\cite{Teu-AdM-91}, the perimeter of $C$ is $\Omega(|S'|)$.

Finally, we bound the perimeter of $C$. In Euclidean geometry, the perimeter of a set $C$ formed by expanding a hull $H$ by distance $\epsilon/2$ can be calculated exactly, as the perimeter of $H$ plus that of a ball of radius $\epsilon/2$ (additivity of perimeters for Minkowski sums). In hyperbolic geometry, things are more complicated, but we may still relate the perimeters of $C$ and $H$ as follows.  Partition the perimeter of $H$ into curves of length $O(1)$, and assign each point of the perimeter of $C$ to the nearest such curve of $H$ (Figure~\ref{fig:msthull}, bottom right).  Each curve is assigned only to points within distance $\epsilon/2$ of it, so the total assigned length to each curve is $O(1)$. Therefore, $|H|=\Omega(|C|)$.
Putting these steps together, $|T|=O(|T'|)=O(|S'|)=O(|C|)=O(|H|)$.
\end{proof}

\begin{figure}[t]
\centering
\includegraphics[width=4.5in]{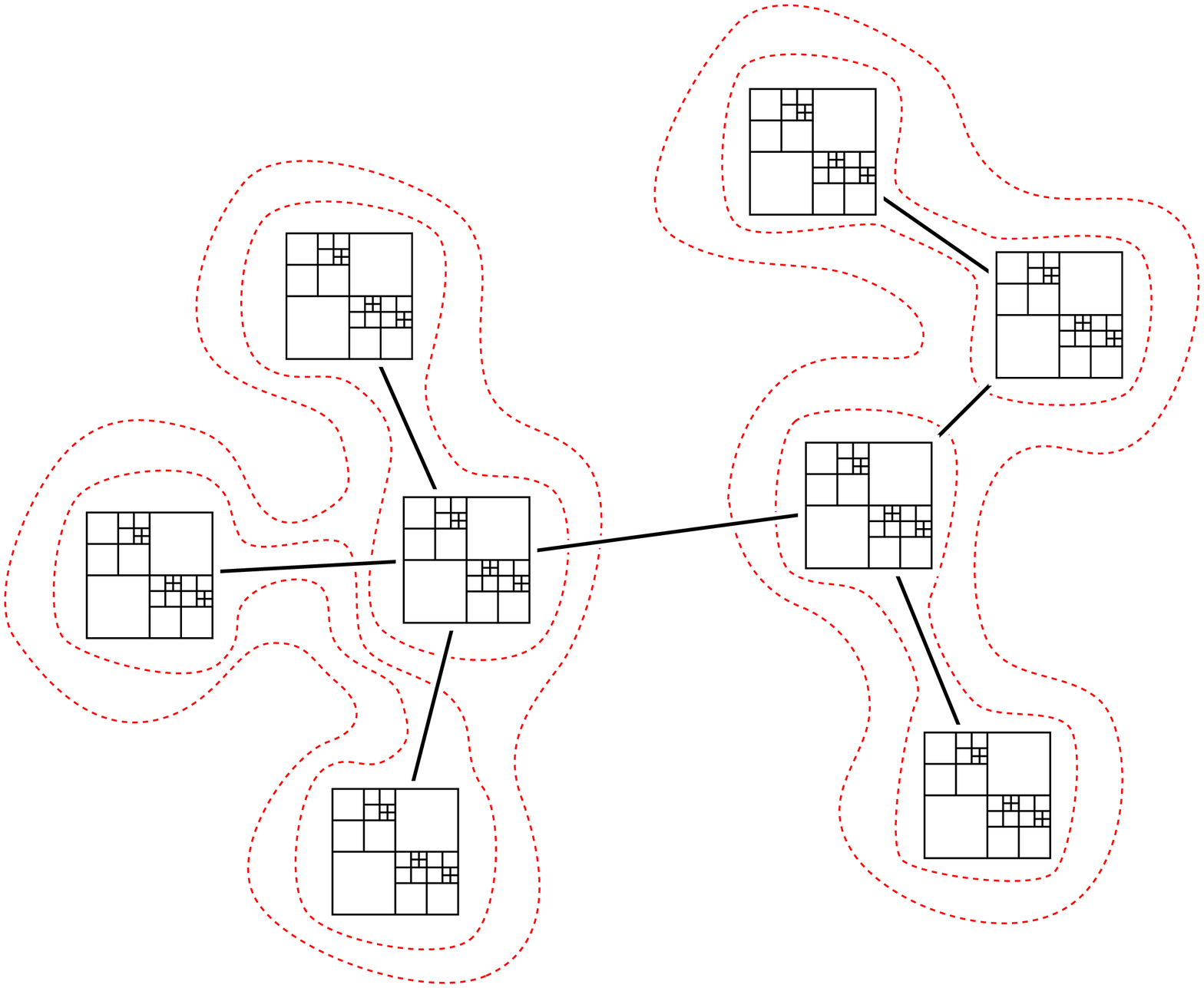}
\caption{Squarepants in a tree: a schematic view of our hyperbolic clustering. We group sites into bounded-diameter subsets, apply our Euclidean quadtree pants decomposition method to each subset, and connect the subsets with clusters following minimum spanning tree edges from our general metric space clustering algorithm.}
\label{fig:sqpintree}
\end{figure}

Thus, if all pairwise distances in our point set are $\Omega(1)$, our tree-based approximation for general metric spaces provides also a constant factor approximation to minimizing the sum of cluster hull perimeters in the hyperbolic plane.  On the other hand, if all pairwise distances are $O(1)$, the Klein model of the hyperbolic plane (in which hyperbolic points are modeled as points in a Euclidean disk, and hyperbolic lines are modeled by straight chords of the disk) provides a convexity-preserving Euclidean approximation of the point set, and applying our quadtree-based approximation to this model leads to a constant factor approximation to the minimum sum of cluster hull perimeters in the hyperbolic plane.  A more challenging situation arises when we are asked to cluster a point set that combines both large and small distances; we show below how to combine these two clustering approaches in an approximation algorithm for this more general case.

Our hyperbolic clustering algorithm, in rough outline, follows the following steps.

\begin{enumerate}
\item Find a subset $S'$ of the input sites, such that $S'$ has no two sites within some constant distance bound of each other, and maximal with respect to this constraint. Group the remaining sites into clusters according to the nearest member of $S'$ to each site.
\item Within each cluster, approximate the sites by a point set in the Euclidean plane, and apply our quadtree-based algorithm to find an approximately-optimal clustering for these points.
\item Find the minimum spanning tree of $S'$, and use our tree-based clustering algorithm for general metric spaces to find a clustering for $S'$. Adjust the boundaries of this clustering so that they surround the smaller clusters associated with each site of $S'$.
\end{enumerate}

Figure~\ref{fig:sqpintree} gives a schematic view of the clustering produced by this algorithm: a collection of quadtrees, connected by clusters that follow minimum spanning tree edges.

\subsection{Finding a Well-Separated Subset}

\begin{figure}[t]
\centering
\includegraphics[width=3in]{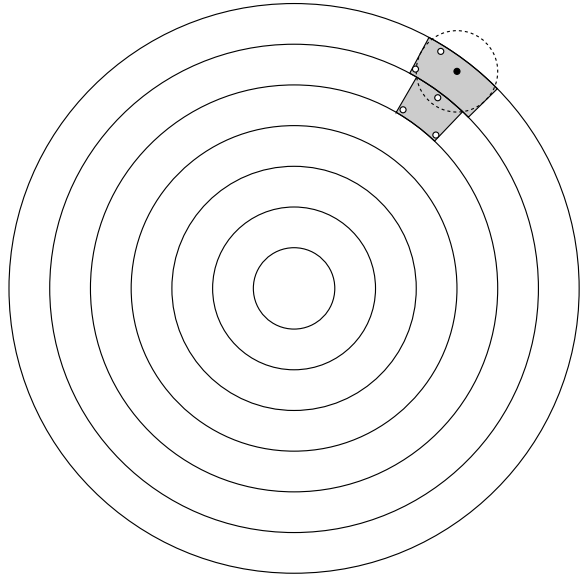}
\caption{Illustration of the algorithm in Lemma~\ref{lem:sparsify} for finding a well-separated subset of the input sites. For each site $p$ that we consider (for instance the site marked by a dark disk) we need only compare its distances to the already-chosen sites (hollow disks) within a bounded radius region $R_{i,p}$ in the same annulus and $R_{i-1,p}$ in the next inner annulus (both shaded).}
\label{fig:sparsify}
\end{figure}

We begin the detailed description of our hyperbolic clustering algorithm by showing how to implement efficiently its first step, in which we find a well-separated subset of the input sites. It is tempting to apply existing Euclidean methods for similar problems~\cite{FedGre-STOC-88}, but these appear to be based on recursive subdivision of rectangles, a concept that makes little sense hyperbolically.

\begin{lemma}
\label{lem:sparsify}
If we are given as input a set $S$ of $n$ point sites in $\H^2$, and a constant $\delta$, we can find in time $O(n\log n)$ a subset $S'\subset S$ such that the closest pair of points in $S'$ are at least at distance $\delta$ apart from each other, and such that every point of $S$ is within distance $\delta$ of some point in $S'$. In the same time bound we may also find the nearest point in $S'$ to each point in $S$, and list all pairs of points of $S'$ with distance less than $2\delta$ of each other.
\end{lemma}

\begin{proof}
We choose an arbitrary point as the origin of our hyperbolic plane, sort the sites by their distance to the origin, and group the sites into subsets, where $S_i$ consists of the sites with distance between $i\delta$ and $(i+1)\delta$ from the origin. Let $A_i$ denote the annulus containing $S_i$, bounded by circles with radii $i\delta$ and $(i+1)\delta$; the circles in Figure~\ref{fig:sparsify} depict the boundaries of these annuli.

Within each annulus $A_i$, we sort the sites in $S_i$ again, in clockwise order by the angle they form with the origin. We will consider the points in this order, adding each site to $S'$ exactly when it is at least $\delta$ in distance from all previously-added sites. In this way, we will generate a maximal subset of sites at distance $\delta$ or more from each other, as desired.

When we consider site $p$, define region $R_{i,p}$ as the smallest annular wedge containing the intersection of $A_i$ with a radius-$\delta$ disk centered at $p$; then only sites within $R_{i,p}\cup R_{i-1,p}$ (the shaded region in Figure~\ref{fig:sparsify}) may be within distance $\delta$ of $p$.
This region has radius at most $2\delta$, as each point in it may be connected by a radial segment of length $\delta$ to the radius-$\delta$ disk; since $\delta$ is assumed to be bounded, and the sites in $S'$ are a bounded distance apart, $|(R_{i,p}\cup R_{i-1,p})\cap S'|=O(1)$.
That is, by scanning sequentially a constant number of steps from $p$ forwards and backwards in the sorted orders for $S'\cap A_i$ and $S'\cap A_{i-1}$, until each sequential search reaches a point outside $R_{i,p}\cup R_{i-1,p}$,
we may find any site in $S'$ that can be within distance $\delta$ of~$p$.
If one of these $O(1)$ candidate neighbors has distance less than $\delta$, we eliminate $p$; otherwise, we add $p$ to $S'$.

\begin{figure}[t]
\centering
\includegraphics[width=3.5in]{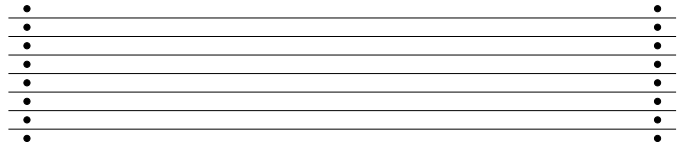}
\caption{Partitioning a set of sites into convex subsets and clustering each subset can have significantly greater total length than a clustering of the overall set.}
\label{fig:badsplit}
\end{figure}

The time for this algorithm is $O(n\log n)$ for the two sorting steps, and then $O(1)$ per point to march stepwise through the sorted orders for $S_i$ and $S'\cap A_{i-1}$ and determine whether to keep or eliminate the point, for a total time of $O(n\log n)$.

Once we have constructed $S'$, finding the nearest neighbor in $S'$ for each point in $S$ may be done by a similar scan through a constant number of sites in $(R_{i,p}\cup R_{i-1,p}\cup R_{i+1,p})\cap S'$,
and listing pairs of sites in $S'$ within distance $2\delta$ of each other may be performed via a similar search of five neighboring annuli for each site.
\end{proof}

We leave as an open problem efficiently solving the generalized version of this problem in which $\delta$ is not restricted to be a constant; the difficulty with this generalization is that the regions $R_{i,p}$ as defined in the proof above may contain nonconstant numbers of sites of $S'$.

\subsection{Clustering Low-Diameter Neighborhoods}

Form the Voronoi diagram of the sites in the subset $S'$ described in Lemma~\ref{lem:sparsify}. For any site $c_i\in S'$, let $V_i$ be formed as the intersection of the Voronoi cell of $c_i$ with a disk of radius $\delta$ centered on $c_i$.  Each cell $V_i$ is a convex and has bounded diameter.  Therefore, an optimal clustering for the sites in $V_i$ may be approximated easily enough, as follows.

Embed $V_i$ in a Klein model of the hyperbolic plane; that is, a unit disk in the Euclidean plane, in which hyperbolic lines are modeled as straight line segments through the disk. This model has the convenient property that convex hulls in the hyperbolic plane are modeled by convex hulls of the corresponding points in the model. Choose the embedding in such a way that $c_i$ is placed at the center of the disk. As the sites in $V_i$ have bounded distance from the center, the distortion of the embedding is also bounded. Then, use our quadtree based algorithm to approximate the optimal clustering of the embedded sites. By the bound on the distortion, the result is also an approximation to the optimal clustering of the same sites in the original hyperbolic plane.

However, we need a stronger conclusion. It is not enough that each cell $V_i$ is clustered approximately optimally. Rather, we need the sum of the cluster lengths in all cells to approximate the optimal clustering of the entire set. To see that this is a nontrivial requirement, see Figure~\ref{fig:badsplit}: there exists a point set, and a partition of the point set into convex regions, such that the sum of the optimal clustering lengths within each region is much larger than the optimal clustering of the overall point set. However, as we now show, such a phenomenon can not occur with our partition into Voronoi cells due to the low aspect ratio of these cells.

\begin{figure}[t]
\centering
\includegraphics[width=4.5in]{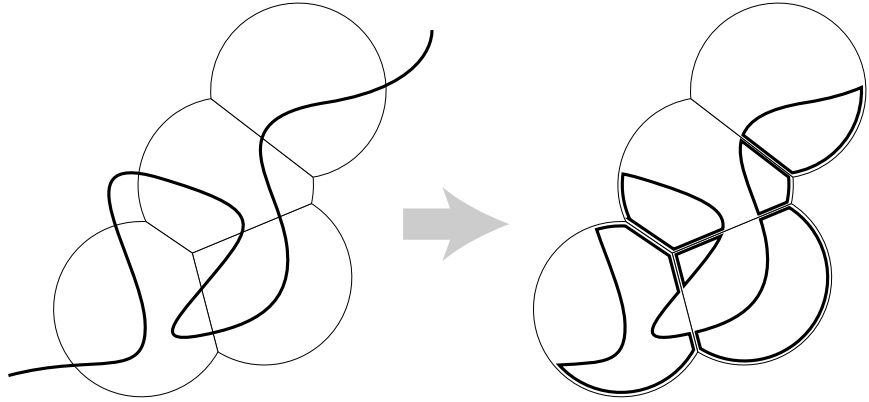}
\caption{Any curve passing through the cells $V_i$ can be divided into components in each cell, increasing the total length by at most a constant factor (Lemma~\ref{lem:curvecells}).}
\label{fig:curvecells}
\end{figure}

\begin{figure}[t]
\centering
\includegraphics[width=2in]{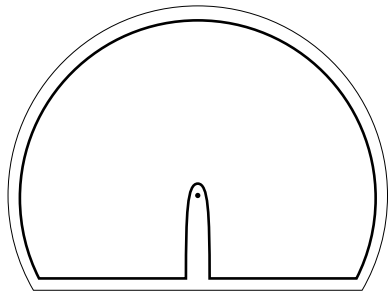}
\caption{Worst-case configuration for Lemma~\ref{lem:curvecells}: a path of length $\delta$ connects two nearby points on the boundary, avoiding $c_i$, so when forming $C_i$ we must use nearly all of the boundary of cell $V_i$.}
\label{fig:ccwc}
\end{figure}

\begin{lemma}
\label{lem:curvecells}
Let $C$ be a simple closed curve, and $T$ be the sites inside $C$. Then there exists a collection of  curves $C_i$, each consisting of a disjoint union of zero or more simple closed curves, such that each $C_i$ encloses $T\cap V_i$, and such that the total length of all the curves $C_i$ is proportional to a constant times the length of $C$.
\end{lemma}

\begin{proof}
We construct the curves $C_i$ by intersecting $C$ with each cell $V_i$, and reconnecting each end of the resulting curves by a curve passing around the boundary of cell $V_i$, as shown in Figure~\ref{fig:curvecells}. 

We must show that the additional length created in this reconnection process is at most proportional to the length of~$C$. We distinguish the components of the resulting collection of curves into two cases: components containing the center $c_i$ of their cell $V_i$, and all other components.

If $C$ contains $k$ centers of cells, and passes through two or more cells, then by Lemma~\ref{lem:msthull} the length of $C$ must be $\Omega(k)$. Within each cell $V_i$, the total length of the connection between the endpoints of the component of $C$ containing $c_i$ is $O(1)$, so the total length in all such cells is $O(k)$. Therefore, the total reconnection length for these components is proportional at most to that of $C$.

Finally, suppose $C_i$ is a component of the collection of curves, not containing $c_i$, and let $r$ be the ratio between the length of the portion of the boundary of $V_i$ bounding $C_i$ and the portion of $C$ bounding $C_i$. The maximum possible value of $r$ occurs when $C$ connects the two points where $C$ crosses the boundary of $V_i$ by as short a path as possible (avoiding $c_i$), and where the portion of boundary between these two crossing points is as long as possible: that is, except possibly for the straight line segments containing the two crossing points, the boundary of $V_i$ follows the radius-$\delta$ disk centered around $c_i$. Next, to maximize $r$,
 the length of the path formed by intersecting $C$ with $V_i$ should be as short as possible. That is, the two endpoints must be at distance exactly $\delta/2$ from $c_i$, as no boundary point of $V_i$ may be closer to $c_i$ than that. Finally, among all ways of choosing two endpoints at distance $\delta/2$ from $c_i$, the one maximizing $r$ is the one where the two endpoints are coincident, and the portion of $C_i$ following the boundary of $V_i$ follows the entire boundary (Figure~\ref{fig:ccwc}). For this worst-case configuration, $r$ is constant whenever $\delta$ is constant. Therefore, the total reconnection length for components of this section type is again proportional at most to that of $C$.
\end{proof}

\begin{corollary}
\label{cor:hquad}
If we are given as input a set of sites in $\H^2$, group them into cells $V_i$ as described above, use the Klein model to map each cell with bounded distortion in the Euclidean plane, and apply our quadtree clustering method to the resulting family of sets of Euclidean sites, then
the total length of the resulting clusterings is at most proportional to the total length of the optimal clustering of all the sites.
\end{corollary}

\begin{proof}
From the optimal clustering of all the points we can form a clustering within each $V_i$ by restricting the clusters to the sites within $V_i$. By applying Lemma~\ref{lem:curvecells} to the convex hulls of each cluster, we see that the total length of the convex hulls of each of these clusterings is at most proportional to the total length of the original optimal clustering. Since the quadtree clustering approximates the optimal clustering within each cell $V_i$, the sum of the quadtree clusterings is again within a constant factor of the overall optimal clustering.
\end{proof}

\subsection{Clusters of Clusters}

We now show how to find larger clusters in our clustering, connecting multiple sites of $S'$. Our algorithm for this part follows that for general metric spaces: we form the minimum spanning tree of $S'$, expand it by zero-length edges so that all internal vertices have degree three, and repeatedly split the resulting tree on the edge that most evenly balances the weights of the two subtrees formed by the split.

\begin{lemma}
\label{lem:hbigclust}
Let $S$, $S'$, and the cells $V_i$ be as described above.
Then in $O(n\log n)$ time, we can find a noncrossing family of curves, including among them the convex hulls of the sites in each cell $V_i$, such that each component of the plane minus these curves is either one of these convex hulls or a pair of pants.
The total length of these curves is proportional to that of the optimal clustering of $S$.
\end{lemma}

\begin{proof}
We note that the optimal clustering of $S$ has length at least that of the optimal clustering of $S'$, as we can omit redundant boundary curves from any clustering of $S$ to form a clustering of $S'$.
Further, by Lemma~\ref{lem:msthull}, the length of the optimal clustering of $S'$ in terms of the sum of convex hulls of clusters is proportional to the length of the optimal clustering in terms of the sum of spanning tree lengths of clusters.

As in our general metric clustering algorithm, we form the minimum spanning tree of $S'$, expand it by zero-length edges so that all internal vertices have degree three, and repeatedly split the resulting tree on the edge that most evenly balances the weights of the two subtrees formed by the split.
In order to reduce crossings between clusters for our application of this method to pants decomposition, whenever we split a vertex, we choose a split consistent with the radial ordering of the neighbors to that vertex, so that the expanded tree could be viewed as embedded without crossings in the hyperbolic plane, with the points into which each original vertex has been expanded placed within a small disk near the original point location.

We then cluster this expanded tree by greedy splitting as in the general metric clustering algorithm.
We surround each cluster of the clustering by a curve that follows the boundary of the hull of a cell $V_i$ whenever the subtree corresponding to that cluster passes through $c_i$, and that (outside these hulls) follows the Euler tour of the minimum spanning tree of $S'$.

If the subtree corresponding to the cluster passes through $k$ hulls, its length is increased by $O(k)$, but (by Lemma~\ref{lem:msthull}) the length of the convex hull of the cluster must already have been $\Omega(k)$.  Therefore, the total length of this system of curves is proportional to that of the clustering of $S'$, which (by Theorem~\ref{thm:metric}) is proportional to that of the optimal clustering.
\end{proof}

\subsection{Hyperbolic Clustering and Pants}

We are now ready to put these results together, in our overall hyperbolic clustering algorithm.

\begin{theorem}
\label{thm:hclust}
In $O(n\log n)$ time, we can construct a hierarchical clustering with total cluster perimeter within a constant factor of optimal for any set of hyperbolic sites.
\end{theorem}

\begin{proof}
We construct a set of $\delta$-separated points by Lemma~\ref{lem:sparsify}, form the Voronoi cells $V_i$ (each such cell being formed as the intersection of a disk with a constant number of halfplanes, the neighbors of each center $c_i$ being found as described in Lemma~\ref{lem:sparsify}), and assign each site to the nearest center $c_i$  as described in Lemma~\ref{lem:sparsify}. We then form clusters within each cell $V_i$ as described in Corollary~\ref{cor:hquad} and clusters grouping sets of cells as described in Lemma~\ref{lem:hbigclust}.
\end{proof}

The same methods and proof apply, essentially unchanged, for hyperbolic pants decomposition.
The quadtree clustering used within each cell $V_i$ can be modified as in the Euclidean case to produce a family of noncrossing curves with essentially the same length, and
Lemma~\ref{lem:hbigclust} already describes how to form noncrossing curves for its clusters.

\begin{theorem}
\label{thm:hpants}
In $O(n\log n)$ time, we can construct a pants decomposition with total length within a constant factor of optimal for any set of hyperbolic sites.
\end{theorem}

\section{Conclusion}

We have provided efficient approximation algorithms for clustering in general metric spaces and the Euclidean and hyperbolic planes. We suspect that our NP-completeness proof for general metric clustering can be extended to show that it is hard to approximate the problem within a factor better than $1+\epsilon$, for some fixed $\epsilon$, but it would nevertheless be of interest to reduce the 3.42 approximation ratio of our algorithm.

For the Euclidean and hyperbolic problems, we do not know whether finding the optimal clustering is intractable. Also, the approximation ratios of our algorithms for these problems are large and inexplicit. It would be of interest to resolve the computational complexity of these problems, and to find improved approximations for them.

Finally, there are many other objective functions for clustering quality than the ones we have considered.  As an example, it is natural to consider clustering planar point sets by convex hull area instead of perimeter; this version of the problem is closely related to covering point sets by few lines \cite{AgaPro-Algs-03,AgaProVar-CGTA-03}, as collinear point sets lead to zero-area clusters. Much work remains to be done on algorithms for optimizing or approximating optimal clusterings for this and other criteria.

\section*{Acknowledgements}

A preliminary version of this paper appeared in the Proceedings of the 18th ACM-SIAM Symposium on Discrete Algorithms, 2007. We thank Mohammadreza Ghodsi for pointing out an error in an earlier version of the proof of Lemma~\ref{lem:E-lb}.

\raggedright
\bibliographystyle{abuser}
\bibliography{squarepants}

\end{document}